\documentclass[12pt]{article}
\usepackage{amssymb,amsmath,epsfig}

\begin{document}

\title{\bf Dynamical Instability and Expansion-free Condition in $f(R, T)$ Gravity }

\author{Ifra Noureen \thanks{ifra.noureen@gmail.com} ${}^{(a)}$, M. Zubair
\thanks{mzubairkk@gmail.com; drmzubair@ciitlahore.edu.pk} ${}^{(b)}$, \\
${}^{(a)}$ Department of Mathematics,\\University of Management and Technology, Lahore, Pakistan. \\
${}^{(b)}$ Department of Mathematics,\\ COMSATS Institute of
Information Technology, Lahore, Pakistan.}
\date{}
\maketitle

\begin{abstract}
Dynamical analysis of spherically symmetric collapsing star
surrounding in locally anisotropic environment with expansion-free
condition is presented in $f(R,T)$ gravity, where $R$ corresponds to
Ricci scalar and $T$ stands for the trace of energy momentum tensor.
The modified field equations and evolution equations are
reconstructed in the framework of $f(R,T)$ gravity. In order to
acquire the collapse equation we implement the perturbation on all
matter variables and dark source components comprising the viable
$f(R,T)$ model. The instability
range is described in Newtonian and post-Newtonian approximation. It is
observed that the unequal stresses and density
profile defines instability rage rather then the adiabatic index. However,
the physical quantities are constrained to maintain positivity of energy density
and stable stellar configuration.
\end{abstract}


\section{Introduction}

The astrophysics and astronomical theories are invigorated largely
by the gravitational collapse and instability range explorations of
self gravitating objects. Celestial objects tend to collapse when
they exhaust all their nuclear fuel, gravity takes over as the
inward governing force. The gravitating bodies undergoing collapse
face contraction to a point that results in high energy dissipation
in the form of heat flux or radiation transport \cite{1}. The end
state of stellar collapse has been studied extensively, a continual
evolution of compact object might end up as a naked singularity or
as black hole depending upon the size of collapsing star and also on
the background that plays important role in pressure to gravity
imbalances \cite{2}-\cite{2b}.

The gravitating objects are interesting only when they are stable
against fluctuations, supermassive stars tends to be more unstable
in comparison to the less massive stars \cite{3}. Instability
problem in star's evolution is of fundamental importance,
Chandrasekhar \cite{4} presented the primary explorations on
dynamical instability of spherical stars. He identified instability
range of star having mass $M$ and radius $r$ by a factor $\Gamma$
pertaining the inequality $\Gamma\geq\frac{4}{3}+n\frac{M}{r}$.
Adiabatic index measures compressibility of the fluid i.e.,
variation of pressure with a given change in the density. The
analysis of expanding and collapsing regions in gravitational
collapse was presented by Sharif and Abbas \cite{ab}.

Herrera and his collaborators \cite{5}-\cite{8} presented the
dynamical analysis associated with isotropy, local anisotropy,
shear, radiation and dissipation with the help of $\Gamma$, it was
established that minor alterations from isotropic profile or slight
change in shearing effects bring drastic changes in range of
instability. However, instability range of stars with zero expansion
does not depend on stiffness of fluid, rather on other physical
parameters \cite{9}-\cite{11}, such as mass distribution, energy
density profile, radial and tangential pressure. The impact of local
anisotropy on plane expansion-free gravitational collapse is studied
in \cite{az}.

General Relativity (GR) facilitates in providing field equations
that leads to the dynamics of universe in accordance with its
material ingredients. The predictions of GR are suitable for small
distances, however, there are some limitations of GR in description
of late time universe. Modified gravity theories have been widely
used to incorporate dark energy components of universe by inducing
alterations in Einstein Hilbert (EH) action. Due to modifications in
laws of gravity at long distances, dark source terms of modified
gravity leaves phenomenal observational signatures such as cosmic
microwave background, weak lensing and galaxy clustering
\cite{12}-\cite{15}. Many people investigated the dynamics of
collapse and instability range in modified theories of gravity,
Cembranos et al. \cite{16} studied the collapse of self-gravitating
dust particles. Sharif and Rani \cite{R} established the instability
range of locally anisotropic non-dissipative evolution in $f(T)$
theory.

Among modified gravity theories, $f(R)$ exhibits the most elementary
modifications to EH action by adopting a general function $f(R)$ of
Ricci scalar. Ghosh and Maharaj \cite{17} indicated that null dust
non-static collapse in $f(R)$ around de-Sitter higher dimensional
background leads to naked singularity. Combined effect of
electromagnetic field and viable $f(R)$ model has been investigated
in \cite{18}, concluding that inclusion of Maxwell source tends to
enhance the stability range. Borisov et al. \cite{19} investigated
the spherically symmetric collapse of $f(R)$ models with non-linear
coupling scalar by execution of one-dimensional numerical
simulations. The dynamical instability of extremal Schwarzschild
de-Sitter background framed in $f(R)$ is investigated in \cite{20}.

Another modification of GR and generalization of $f(R)$ was
presented in 2011 by Harko et al. \cite{21} termed as  $f(R,T)$
gravity theory constituting the matter and geometry coupling, EH
action is modified in a way that gravitational Lagrangian includes
higher order curvature terms alongwith the trace of energy momentum
tensor $T$. Shabani and Farhoudi \cite{22} explained the weak field
limit by applying dynamical system approach and analyzed the
cosmological implications of $f(R,T)$ models with a variety of
cosmological parameters such as Hubble parameter, its inverse, snap
parameters, weight function, deceleration, jerk and equation of
state parameter. Ayuso et al. \cite{22a}
worked on consistency criterion for non-minimally coupled class of modified
theories of gravity. Sharif and Zubair \cite{23}- \cite{25} ascertained
the laws of thermodynamics, energy conditions and analyzed the
anisotropic universe models in $f(R,T)$ framework.

Chakraborty \cite{26} explored various aspects of homogeneous and isotropic
cosmological models in $f(R,T)$ and formulate the energy conditions
for perfect fluid. Dynamics of scalar perturbations in $f(R,T)$ is explored
in \cite{26a}. Jamil et al. \cite{26b, 26c} reconstruct some cosmological models and
studied laws of thermodynamics in $f(R,T)$. In a recent paper \cite{27}, dynamical
instability of isotropic collapsing fluid in the context of $f(R,T)$
is considered. We have also discussed the stability analysis of
spherically symmetric collapsing star surrounding in locally
anisotropic environment in $f(R,T)$ gravity \cite{27a}. Furthermore,
conditions on physical quantities are constructed for Newtonian
and post-Newtonian eras to address instability problem.

Herein, we intend to develop the instability range of $f(R,T)$ model
under anisotropic background constraining to zero expansion. The
expansion-free condition necessarily implies the appearance of
cavity within fluid distribution that might help in modeling of
voids at cosmological scales. Also, such distributions must bear
energy density inhomogeneities that are incorporated here by
inducing non-constant energy density and pressure anisotropy. The
dynamical analysis of various fluid distributions with
expansion-free condition has been studied in $f(R)$
\cite{28}-\cite{30}, however, such situations have not been covered
yet in $f(R,T)$. Recently, Noureen and Zubair \cite{30a} discussed the
implications of extended Starobinsky Model on dynamical instability
of axially symmetric gravitating body.

To develop collapse equation in $f(R,T)$, we construct corresponding
field equations constituting expansion-free fluid. The action in
$f(R,T)$ is as in \cite{21}
\begin{equation}\label{1}
\int dx^4\sqrt{-g}[\frac{f(R, T)}{16\pi G}+\mathcal{L} _ {(m)}],
\end{equation}
where $\mathcal{L} _ {(m)}$ is matter Lagrangian and $g$ denotes the
metric tensor. The Lagrangian $\mathcal{L} _ {(m)}$ can assume
various choices, each choice corresponds to a set of field equations
for some special form of fluid. Here, we have chosen $\mathcal{L} _
{(m)}= \rho$, $8\pi G = 1$ and upon variation of above action with
metric $g_{uv}$ the field equations are formed as
\begin{eqnarray}\nonumber
G_{uv}&=&\frac{1}{f_R}\left[(f_T+1)T^{(m)}_{uv}-\rho g_{uv}f_T+
\frac{f-Rf_R}{2}g_{uv}\right.\\\label{4}&+&\left.(\nabla_u\nabla_v-g_{uv}\Box)f_R\right],
\end{eqnarray}
where $T^{(m)}_{uv}$ denotes the energy momentum tensor for usual
matter.

The matter Lagrangian is configured in a way that it depends only on
the components of metric tensor \cite{31}. In order to present the
dynamical analysis we implement the linear perturbation on collapse
equation, assuming that initially all physical quantities are in
static equilibrium. The paper is arranged as: Einstein's field
equations and dynamical equations for $f(R,T)$ are constructed in
section \textbf{2} that leads to the collapse equation. In section
\textbf{3} perturbation scheme is implemented to the dynamical
equations. Section \textbf{4} covers the discussion of
expansion-free condition and the components affecting the stability
of gravitating objects, extracted from perturbed Bianchi identities
along with corrections to Newtonian and post-Newtonian eras and GR
solution. Section \textbf{5} comprises the summary followed by an
appendix.

\section{Dynamical Equations in $f(R,T)$}

We choose a three dimensional external spherical boundary surface
$\Sigma$ that pertains two regions of spacetime termed as interior
and exterior regions. The line element for region inside the
boundary $\Sigma$ is of the form
\begin{equation}\label{1'}
ds^2_-=W^2(t,r)dt^{2}-X^2(t,r)dr^{2}-Y^2(t,r)(d\theta^{2}+\sin^{2}\theta
d\phi^{2}).
\end{equation}
The domain beyond (lying outside) $\Sigma$ is exterior region with
following line element \cite{30}
\begin{equation}\label{25}
ds^2_+=\left(1-\frac{2M}{r}\right)d\nu^2+2drd\nu-r^2(d\theta^2+\sin^{2}\theta
d\phi^{2}),
\end{equation}
where $\nu$ is the corresponding retarded time and $M$ is total
mass. To arrive at onset of field equations given in Eq.(\ref{4}),
we choose $T^{(m)}_{uv}$ describing anisotropic fluid distribution
of usual matter, given as
\begin{equation}\label{5}
T^{(m)}_{uv}=(\rho+p_{\bot})V_{u}V_{v}-p_{\bot}g_{uv}+(p_{r}-p_{\bot})\chi_{u}\chi_{v},
\end{equation}
where $\rho$ denotes energy density, $V_{u}$ is four-velocity of the
fluid, $\chi_u$ corresponds to radial four vector, $p_r$ and
$p_{\bot}$ represent the radial and tangential pressures,
respectively. The physical quantities appear in energy momentum
tensor are in accordance with the following identities
\begin{equation}\label{3}
V^{u}=W^{-1}\delta^{u}_{0},\quad V^{u}V_{u}=1,\quad
\chi^{u}=X^{-1}\delta^u_1,\quad \chi^{u}\chi_{u}=-1.
\end{equation}
The expansion scalar $\Theta$ defines rate of change of small
volumes of the fluid, given by
\begin{equation}\label{3'}
\Theta=V^u_{;u}=\frac{1}{W}\left(\frac{\dot{X}}{X}+2\frac{\dot{Y}}{Y}\right),
\end{equation}
where dot and prime denote the time and radial derivatives
respectively. The components of field equations for spherically
symmetric interior spacetime are of the form
\begin{eqnarray}\nonumber
G_{00}&=&\frac{1}{f_R}\left[\rho+ \frac{f-Rf_R}{2}+
\frac{f_R''}{X^2}-\frac{\dot
f_R}{W^2}\left(\frac{\dot{X}}{X}+\frac{2\dot{Y}}{Y}\right)\right.\\\label{f1}
&&\left.
-\frac{f_R'}{X^2}\left(\frac{X'}{X}-\frac{2Y'}{Y}\right)\right]
,\\\label{f2} G_{01}&=&\frac{1}{f_R}\left[\dot{f_R}'
-\frac{W'}{W}\dot{f_R}-\frac{\dot{X}}{X}f_R'\right],\\\nonumber
G_{11}&=&\frac{1}{f_R}\left[p_r+\left(\rho+p_r\right)f_T-
\frac{f-Rf_R}{2}+ \frac{\ddot{f_R}}{W^2}-\frac{\dot
f_R}{W^2}\left(\frac{\dot{W}}{W}-\frac{2\dot{Y}}{Y}\right)
\right.\\\label{f3}
&&\left.-\frac{f_R'}{X^2}\left(\frac{W'}{W}+\frac{2Y'}{Y}\right)\right],
\\\nonumber
G_{22}&=&\frac{1}{f_R}\left[p_\perp+\left(\rho + p_\perp\right)f_T-
\frac{f-Rf_R}{2}+
\frac{\ddot{f_R}}{W^2}-\frac{f_R''}{X^2}-\frac{\dot
f_R}{W^2}\left(\frac{\dot{W}}{W} \right.\right.\\\label{f4}
&&\left.\left.-\frac{\dot{X}}{X}-\frac{\dot{Y}}{Y}\right)
-\frac{f_R'}{X^2}\left(\frac{W'}{W}-\frac{X'}{X}+\frac{Y'}{Y}\right)\right].
\end{eqnarray}
The dynamical equations are important in establishment of the
instability range of collapsing stars. Misner and Sharp mass
function furnishes the total amount of energy in a spherical star of
radius "Y" and facilitates in formulation of dynamical equations,
given by \cite{32}
\begin{equation}\label{3''}
m(t,r)=\frac{Y}{2}\left(1+\frac{\dot{Y}^2}{W^2}-\frac{Y'^2}{B^2}\right),
\end{equation}
The matching conditions of adiabatic sphere on the boundary surface
of exterior spacetime results from continuity of differential forms
as \cite{33}
\begin{equation}\label{3'''}
M\overset\Sigma=m(t,r)
\end{equation}
The dynamical analysis can be established by using conservation
laws, we have taken conservation of Einstein tensor because the
energy momentum tensor bear non-vanishing divergence in $f(R,T)$
gravity. The contracted Bianchi identities imply dynamical equations
that further leads to collapse equation, given by
\begin{eqnarray}\label{bb}
&&G^{uv}_{;v}V_{u}=0,\quad
G^{uv}_{;v}\chi_{u}=0,
\end{eqnarray}
Bianchi identities in account with the Eq.(\ref{3'}) become
\begin{eqnarray}
\nonumber &&\dot{\rho}+\rho\left\{[1+f_T]\Theta-\frac{\dot{f_R}}{f_R}\right\}
+[1+f_T]\left\{p_r\frac{\dot{X}}{X}
+2p_\perp\frac{\dot{Y}}{Y}\right\} \\\label{B1}&&+
Z_1(r,t)=0,
\\\nonumber &&
(\rho +p_r)f'_T+(1+f_T)\left\{p'_r+\rho \frac{W'}{W}+
p_r\left(\frac{W'}{W}+2\frac{Y'}{Y}-\frac{f'_R}{f_R}\right)
\right.\\\label{B2}&&\left.-2p_\perp\frac{Y'}{Y}\right\}
+f_T\left(\rho'-\frac{f'_R}{f_R}\right)+Z_2(r,t)=0,
\end{eqnarray}
where $Z_1(r,t)$ and $Z_2(r,t)$ are the corresponding terms
including dark matter components provided in \textbf{Appendix} as
Eqs.(\ref{B3}) and (\ref{B4}) respectively. These equations are
useful in the description of variation from equilibrium to the
evolution.

\section{$f(R,T)$ Model and Perturbation Scheme}

The $f(R, T)$ model we have considered for evolution analysis is
\begin{eqnarray}\label{m}
&&f(R, T)= R+\alpha R^2+\lambda T,
\end{eqnarray}
where $\alpha$ and $\lambda$ corresponds to the positive real
values. Generally, a viable model represents the choice of
parameters whose variation shall be in accordance with the
observational situations \cite{33}. Astrophysical models are
selected by checking their cosmological viability, that must be
fulfilled to extract consistent matter domination phase, to assemble
solar system tests and stable high-curvature configuration
recovering the standard GR. The model under consideration is
consistent with the stable stellar configuration because second
order derivative with respect to $R$ remains positive for assumed
choice of parameters.

The field equations in $f(R, T)$ are highly complicated, their
general solution is a heavier task and has not been accomplished
yet. Evolution of linear perturbations can always be used to study
the gravitational modifications by avoiding such discrepancies. The
concerned collapse equation can be furnished by application of
linear perturbation on dynamical equations along with the static
configuration of field equations leading to the instability range.
The dynamical analysis can be anticipated either by following fixed
or co-moving coordinates i.e., Eulerian or Lagrangian approach
respectively \cite{33}. Since universe is almost homogeneous at
large scale structures that is why we have used co-moving
coordinates.

Initially all physical quantities are considered to be in static
equilibrium so that with the passage of time these have both the
time and radial dependence. Taking $0<\varepsilon\ll1$ the perturbed
form of quantities along with their initial form can be written as
\begin{eqnarray}\label{41} W(t,r)&=&W_0(r)+\varepsilon
D(t)w(r),\\\label{42} X(t,r)&=&X_0(r)+\varepsilon
D(t)x(r),\\\label{43} Y(t,r)&=&Y_0(r)+\varepsilon
D(t)\bar{y}(r),\\\label{44} \rho(t,r)&=&\rho_0(r)+\varepsilon
{\bar{\rho}}(t,r),\\\label{45} p_r(t,r)&=&p_{r0}(r)+\varepsilon
{\bar{p_r}}(t,r),
\\\label{46}
p_\perp(t,r)&=&p_{\perp0}(r)+\varepsilon {\bar{p_\perp}}(t,r),
\\\label{47} m(t,r)&=&m_0(r)+\varepsilon {\bar{m}}(t,r),
\\\label{49'} R(t,r)&=&R_0(r)+\varepsilon D(t)e_1(r),\\\label{50'}
T(t,r)&=&T_0(r)+\varepsilon D(t)e_2(r),\\\nonumber f(R,
T)&=&[R_0(r)+\alpha R_0^2(r)+\lambda T_0]+\varepsilon D(t)e_1(r)
[1\\\label{51'}&+&2\alpha  R_0(r)]+\eta D(t)e_2(r),\\\label{52'}
f_R&=&1+2\alpha R_0(r)+\varepsilon 2\alpha D(t)e_1(r),\\\label{52'}
f_T&=&\lambda, \\\label{52'} \Theta(t,r)&=&\varepsilon\bar{\Theta}.
\end{eqnarray}
Without loss of generality, we have taken the Schwarzschild
coordinate $Y_0(r)=r$ and apply perturbation scheme on dynamical
equations i.e., Eqs.(\ref{B1}) and (\ref{B2}), the perturbed Bianchi
identities turn out to be
\begin{eqnarray}\nonumber&&
\dot{\bar{\rho}}+\left[\frac{2e\rho_0}{1+2\alpha R_0}
+\lambda_1\left\{\frac{2\bar{y}}{r}(\rho_0+2p_{\perp0})
+\frac{x}{X_0}(\rho_0+p_{r0})\right\}\right.\\\label{B1p}&&\left.
+(1+2\alpha R_0)Z_{1p}\right]\dot{D}=0,
\\\nonumber
&&\lambda_1\left\{\bar{p_r}'+\bar{\rho}\frac{W'_0}{W_0}
+\bar{p_r}\left(\frac{W'_0}{W_0}+\frac{2}{r}-\frac{2\alpha R'_0}{1
+2\alpha
R_0}\right)-\frac{2\bar{p_\perp}}{r}\right\}+2\alpha\left[\frac{1}{W_0^2}\left(e'
\right.\right.\\\nonumber &&\left.\left.
+2e\frac{X'_0}{X_0}-\frac{x}{X_0}R'_0\right)+X_0^2(1 +2\alpha
R_0)\left\{\frac{e}{X_0^2(1+2\alpha R_0)}\right\}'\right]\ddot{D}
+D\left[\lambda_1[(\rho_0\right.
\end{eqnarray}
\begin{eqnarray}\nonumber
&&\left.+p_{r0})(\frac{w}{W_0})'
-2(p_{r0}+p_{\perp0})(\frac{\bar{y}}{r})']+\lambda\bar{\rho'}
-\frac{2\alpha}{1+2\alpha R_0}\left\{\lambda_1\left(p'_{r0}
+\rho_0\frac{W'_0}{W_0}
\right.\right.\right.\\\nonumber&&\left.\left.\left.
+p_{r0}\left(\frac{2}{r}+\frac{W'_0}{W_0}-\frac{2\alpha R'_0}{1
+2\alpha R_0}\right)\right)\right\}
+\lambda\left(e'+e[\rho'_0-\frac{2\alpha R'_0}{1+2\alpha
R_0}]\right) \right.\\\label{B2p}&&\left.+(1+2\alpha
R_0)Z_{2p}\right]=0.
\end{eqnarray}
For the sake of simplicity we assume that $e_1=e_2=e$ and set
$\lambda_1=\lambda+1$, $Z_{1p}$ and $Z_{2p}$ are provided in
appendix.

Elimination of $\dot{\bar{\rho}}$ from Eq. (\ref{B1p}) and
integration of resultant with respect to time yields an expression
for $\bar{\rho}$ of following form
\begin{equation}\label{B1p'}
\bar{\rho}=-\left[\frac{2e\rho_0}{1+2\alpha R_0}+\lambda_1\left\{\frac{2\bar{y}}{r}(\rho_0
+2p_{\perp0})+\frac{x}{X_0}(\rho_0+p_{r0})\right\}+(1+2\alpha R_0)Z_{1p}\right]D.
\end{equation}
The perturbed on field equation Eq.(\ref{f4}) leads to the
expression for $\bar{p}_\perp$, that turns out to be
\begin{eqnarray}\nonumber &&
\bar{p}_\perp=\frac{\ddot{D}}{W_0^2}\left\{\frac{(1+2\alpha R_0)\bar{y}}{r}-2\alpha e\right\}
-\frac{\lambda \bar{\rho}}{\lambda_1}+\left\{\left(p_{\perp0}
-\frac{\lambda}{\lambda_1}\rho_0\right)\frac{2\alpha e}{1+2\alpha R_0}\right.
\\\label{B9}&&\left.+\frac{Z_3}{\lambda_1}\right\}D,
\end{eqnarray}
effective part of the field equation is denoted by $Z_3$, given in appendix
Eq.(\ref{Z3}).
Matching conditions at boundary surface reveals
\begin{equation}\label{3'''}
p_r\overset\Sigma=0, \quad p_\perp\overset\Sigma=0
\end{equation}
Above equation together with perturbed form of Eq.(\ref{f4}) can be
written in the following form
\begin{equation}\label{66}
\ddot{D}(t)-Z_4(r) D(t)=0,
\end{equation}
provided that
\begin{eqnarray}\nonumber
Z_4&=&\frac{rW_0^2}{(1+2\alpha R_0)\bar{y}-2\alpha
er}\left[\frac{2\alpha e}{1 +2\alpha
R_0}p_{\perp0}+\lambda\left\{\frac{2\bar{y}}{r}(\rho_0
+2p_{\perp0})\right.\right.\\\label{67}&&\left.\left.
+\frac{x}{X_0}(\rho_0+p_{r0})\right\}+(1+2\alpha
R_0)Z_{1p}+\frac{Z_3}{\lambda_1}\right].
\end{eqnarray}
The solution of Eq.(\ref{66}) takes form
\begin{equation}\label{68}
D(t)=-e^{\sqrt{Z_4}t}.
\end{equation}
The terms appearing in $Z_4$ are presumed in a way that all terms remain positive to have
a valid solution for $D$.
Expansion-free condition and stability range is discussed
in the following section.

\section{Expansion-free Condition with Newtonian and Post-Newtonian Limits}

The models with an additional zero expansion condition delimitate
two hypersurfaces, one separates the external Schwarzschild solution
from the fluid distribution and other is the boundary between fluid
distribution and internal cavity. Such models have extensive
astrophysical applications where cavity within fluid distribution
exists and are significant in investigation of voids at cosmological
scales \cite{35}. The spongelike structures are termed as voids
existing in different sizes i.e., mini-voids to super-voids
\cite{36, 37} accompanying almost $50\%$ of the universe, considered
as vacuum spherical cavities within fluid distribution.

Implementation of linear perturbation on Eq.(\ref{3'}) and
Eq.(\ref{3''}) respectively, implies
\begin{eqnarray}\label{01} &&\bar{\Theta}=\frac{\dot{D}}{W_0}
\left(\frac{x}{X_0}+2\frac{\bar{y}}{r}\right)'
\\\label{02} &&m_0=\frac{r}{2}\left(1-\frac{1}{X_0^2}\right),
\\\label{03} &&\bar{m}=-\frac{D}{X_0^2}\left[r\left(\bar{y}'-\frac{x}{X_0}\right)
+(1-X_0^2)\frac{\bar{y}}{2}\right].
\end{eqnarray}
The expansion free condition implies vanishing expansion scalar
i.e., $\Theta=0$ implying
\begin{equation}\label{04}
\frac{x}{X_0}=-2\frac{\bar{y}}{r}.
\end{equation}
In order to present the dynamical analysis in Newtonian (N) and
post-Newtonian (pN) limits, we assume
\begin{equation}\label{05}
\rho_0\gg p_{r0}, \quad \rho_0\gg p_{\perp0}.
\end{equation}
The metric coefficients upto the pN approximation in c.g.s. units
are taken as
\begin{equation}\label{06}
W_0=1-\frac{Gm_0}{c^2r}, \quad X_0=1+\frac{Gm_0}{c^2r},
\end{equation}
where $c$ denote speed of light and $G$ stands for gravitational
constant. Expression for $\frac{X'_0}{X_0}$ can be obtained from Eq.
(\ref{02}) as
\begin{equation}\label{08}
\frac{X'_0}{X_0}= \frac{-m_0}{r(r-2m_0)},
\end{equation}
Eq. (\ref{02}) together with (\ref{f3}) implies
\begin{eqnarray}\nonumber
\frac{W'_0}{W_0}&=&\frac{1}{2r(r-2m_0)(1+2\alpha R_0+r\alpha
R_0')}\left[r^3(\lambda_1
p_{r0}+\lambda\rho_0-R_0\right.\\\label{09}&&\left. -3 \alpha
R_0^2)+2\alpha r(R_0-2rR_0'+4R_0'm_0)+2m_0\right].
\end{eqnarray}
The static configuration of first Bianchi identity is identically
satisfied while second provides a fruitful result in terms of
dynamical equation. Substitution of Eqs. (\ref{08}) and (\ref{09})
in statically configured Eq. (\ref{B2}) and after some manipulation
the dynamical equation in relativistic units yield
\begin{eqnarray}\nonumber
p'_{r0}&=&-\left[\frac{\lambda}{\lambda_1}\rho_0'+\frac{r(1+2\alpha
R_0)}{r-2m_0}\left[\frac{r-2m_0}{r(1+2\alpha
R_0)}\left\{\frac{\alpha R_0^2}{2}-\frac{2\alpha
R'_0(r-2m_0)}{r}\left(\frac{2}{r}\right.\right.\right.\right.\\\nonumber
&&\left.\left.\left.\left.+\frac{1}{2r(r-2m_0)(1+2\alpha R_0+r\alpha
R_0')}\left[r^3(\lambda_1 p_{r0}+\lambda\rho_0-R_0-3 \alpha
R_0^2)\right.\right.\right.\right.\right.\\\nonumber&&\left.\left.\left.\left.\left.
+2\alpha
r(R_0-2rR_0'+4R_0'm_0)+2m_0\right]\right)\right\}\right]_{,1}
+\frac{2m_0}{r(r-2m_0)}\left(p_{r0}-p_{\perp0}\right.\right.\\\nonumber
&&\left.\left.+\frac{\alpha
R_0^2}{4}-\frac{3}{r}\right)+\frac{2}{r^2}-\frac{2\alpha
R'_0}{1+2\alpha
R_0}\left(p_{r0}+\frac{\lambda}{\lambda_1}\right)+\frac{2\alpha
R''_0(r-2m_0)}{r^2}\right.\\\nonumber
&&\left.+\frac{1}{2r(r-2m_0)(1+2\alpha R_0+r\alpha
R_0')}\left[r^3(\lambda_1 p_{r0}+\lambda\rho_0-R_0-3 \alpha
R_0^2)\right.\right.\\\nonumber&&\left.\left. +2\alpha
r(R_0-2rR_0'+4R_0'm_0)+2m_0\right]\left[\rho_0+p_{r0}+\frac{2\alpha
R'_0(r-2m_0)}{r}\right.\right.\\\nonumber&&\left.\left.
\times\left\{\frac{3m_0}{r(r-2m_0)}-\frac{1}{2r(r-2m_0)(1+2\alpha
R_0+r\alpha R_0')}\left[r^3(\lambda_1
p_{r0}+\lambda\rho_0\right.\right.\right.\right.\\\label{010}&&\left.\left.\left.\left.
-3 \alpha R_0^2)+2\alpha
r(R_0-2rR_0'+4R_0'm_0)+2m_0-R_0\right]\right\}\right]\right].
\end{eqnarray}
In c.g.s. units, we may write above equation as
\begin{eqnarray}\nonumber
p'_{r0}&=&-\left[\frac{c^{-2}r(1+2\alpha
R_0)}{r-2Gc^{-2}m_0}\left[\frac{r-2Gc^{-2}m_0}{rc^{-2}(1+2\alpha
R_0)}\left\{-\frac{2\alpha
R'_0(r-2Gc^{-2}m_0)}{rc^{-2}}\left(\frac{2}{r}\right.\right.\right.\right.\\\nonumber
&&\left.\left.\left.\left.+\frac{1}{2r(r-2c^{-2}m_0)(1+2\alpha R_0+r\alpha
R_0')}\left[r^3(\lambda_1 c^{-2}p_{r0}+\lambda\rho_0-R_0\right.\right.\right.\right.\right.
\\\nonumber&&\left.\left.\left.\left.\left.
-3 \alpha
R_0^2)+2\alpha
r(R_0-2rR_0'+4R_0'Gc^{-2}m_0)+2Gc^{-2}m_0\right]+\frac{\alpha R_0^2}{2}\right)\right\}\right]_{,1}
\right.\\\nonumber
&&\left.+\frac{2Gc^{-2}m_0}{rc^{-2}(r-2Gc^{-2}m_0)}
\left(p_{r0}-p_{\perp0}+\frac{\alpha
R_0^2}{4}-\frac{3}{r}\right)-\frac{2\alpha
R'_0}{1+2\alpha
R_0}\left(c^{-2}p_{r0}
\right.\right.\\\nonumber
&&\left.\left.+\frac{\lambda}{\lambda_1}\right)+\frac{2}{c^{-2}r^2}
+\frac{1}{2r(r-2Gc^{-2}m_0)(1+2\alpha R_0+r\alpha
R_0')}\left[r^3(\lambda_1c^{-2} p_{r0}
\right.\right.\\\nonumber&&\left.\left.+\lambda\rho_0-R_0-3 \alpha
R_0^2) +2\alpha
r(2Gc^{-2}m_0-2rc^{-2}R_0'+4R_0'Gc^{-2}m_0)\right.\right.\\\nonumber&&\left.\left.
+R_0\right]\left[\rho_0+p_{r0}+\left\{\frac{-1}{2r(r-2Gc^{-2}m_0)(1+2\alpha
R_0+rc^{-2}\alpha R_0')}\left[r^3(\lambda\rho_0\right.\right.\right.\right.
\\\nonumber&&\left.\left.\left.\left.
+\lambda_1
c^{-2}p_{r0}-3 \alpha R_0^2)+2\alpha
rc^{-2}(R_0-2rR_0'+4R_0'Gc^{-2}m_0)-R_0
\right.\right.\right.\right.\\\nonumber&&\left.\left.\left.\left.+2Gc^{-2}m_0\right]
+\frac{3Gc^{-2}m_0}{r(r-2Gc^{-2}m_0)}\right\}\frac{2\alpha
R'_0(r-2Gc^{-2}m_0)}{rc^{-2}}\right]+\frac{\lambda}{\lambda_1}\rho_0'
\right.\\\label{011}&&\left.+\frac{2\alpha
R''_0(r-2Gc^{-2}m_0)}{r^2}\right].
\end{eqnarray}
The terms of order $c^0$ and $c^{-2}$ belongs to N and
pN-approximation respectively. One can expand Eq. (\ref{011}) upto
$c^{-2}$ and separate the terms of N and pN limits to distinguish
the physical quantities lying in various regimes.

The use of expansion-free condition in Eq. (\ref{B1p'}) modifies
$\bar{\rho}$ to following form
\begin{equation}\label{012}
\bar{\rho}=-\left[\frac{2e\rho_0}{1+2\alpha R_0}+\lambda_1\frac{2\bar{y}}{r}\left((p_{r0}-
p_{\perp0})\right)+(1+2\alpha R_0)Z_{1p}\right]D.
\end{equation}
The Harrison-Wheeler type equation of state describing second law of
thermodynamics relates $\bar{\rho}$ and $\bar{p_r}$ in terms of
adiabatic index $\Gamma$ as
\begin{equation}\label{013}
\bar{p}_r=\Gamma\frac{p_{r0}}{\rho_0+p_{r0}}\bar{\rho}.
\end{equation}
$\Gamma$ measures the fluid's compressibility belonging to its
stiffness. Inserting $\bar{\rho}$ from Eq.(\ref{012}) in
Eq.(\ref{013}), we have
\begin{equation}\label{013'}
\bar{p}_r=-\Gamma\frac{p_{r0}}{\rho_0+p_{r0}}\left[\frac{2e\rho_0}{1+2\alpha R_0}
+\lambda_1\frac{2\bar{y}}{r}\left((p_{r0}-
p_{\perp0})\right)+(1+2\alpha R_0)Z_{1p}\right]D.
\end{equation}

In view of dimensional analysis, it is found that terms of
$\bar{p}_r$ and $\rho_0\frac{W'_0}{W_0}$ lies in post post Newtonian
(ppN) era and thus can be ignored from the terms of N and pN
approximation. Since we are going to exclude $\bar{p}_r$, so it is
intuitively clear that instability range is independent of $\Gamma$,
no compression is introduced. Use of expansion-free condition
together with the expression found for $D$, it follows that
\begin{eqnarray}\nonumber
&&2\alpha Z_4\left[\frac{1}{W_0^2}\left(e'
+2e\frac{X'_0}{X_0}+2\frac{\bar{y}}{r}R'_0\right)+X_0^2(1
+2\alpha R_0)\left\{\frac{e}{X_0^2(1+2\alpha R_0)}\right\}'\right]
\\\nonumber &&
+\lambda_1[\rho_0(\frac{w}{W_0})'
-2(p_{r0}+p_{\perp0})(\frac{\bar{y}}{r})']+\lambda\bar{\rho'}
-\frac{2\alpha}{1+2\alpha
R_0}\left\{\lambda_1\left(p'_{r0}+p_{r0}\left(\frac{2}{r}
\right.\right.\right.\\\nonumber&&\left.\left.\left. -\frac{2\alpha
R'_0}{1 +2\alpha R_0}\right)\right)\right\}
+\lambda\left(e'+e[\rho'_0-\frac{2\alpha R'_0}{1+2\alpha
R_0}]\right) +(1+2\alpha R_0)Z_{2p}=0,\\\label{014}.
\end{eqnarray}
For simplification of above expression, we take relativistic units
and assume that $\rho_0\gg p_{r0}$, $\rho_0\gg p_{\perp0}$. On
substitution of expressions for $Z_4, W_0, X_0, \frac{X'_0}{X_0}$
and $Z_2p$ respectively from Eqs. (\ref{67}), (\ref{06}), (\ref{08})
and (\ref{Z2p}) yields a very lengthy expression defining the
factors affecting the instability range at N and pN limits. The
expanded version of Eq. (\ref{014}) is large enough, therefore we
are quoting only the results obtained from the collapse equation
together with the restrictions to be imposed on physical parameters.
It is clear from Eq. (\ref{010}) that $p'_{r0}<0$, provided that all
the terms maintain positivity to fulfill stability criterion.
Negative values of $p'_{r0}$ depicts decrease in pressure with time
transition, leading to collapse of gravitating star. Furthermore,
using the c.g.s. units, it is found that the terms of $\bar{p_{r}}$
and $\rho_0\frac{W'_0}{W_0}$ does not take part in evolution for N
and pN approximations since these terms belong to ppN limit. The
analysis of terms lying in N and pN limits imply few restrictions to
be imposed on physical quantities for discussion of instability
range. These are listed as follows:

\begin{itemize}
\item \textbf{Newtonian Regime}: The constraints on material parameters are
\begin{equation}\nonumber
p_{r0}>p_{\perp0}, \quad \alpha^2 r R_0'<1+2\alpha R_0, \quad
\frac{2\alpha R'_0}{1+2\alpha R_0}>\rho_0'-e'.
\end{equation}
The gravitating body remain unstable as long as the inequalities
hold in N approximation.
\item \textbf{post-Newtonian Regime}: In pN limits following restrictions
are found to execute the instability range
\begin{eqnarray}\nonumber&&
p_{r0}>p_{\perp0},\quad r>2m_0, \quad \frac{r}{r+m_0}(xR'_0
+\frac{2em_0}{r})<e'+\lambda,\\\nonumber&& 2\alpha e-(1+2\alpha
R_0)\frac{\bar{y}}{r}>\frac{(r^2-m_0^2)^2}
{r^4}\left\{\frac{er^2}{(r+m_0)}\right\}', \quad
(r-2m_0)>\frac{R_0'}{2R_0}.
\end{eqnarray}
\end{itemize}

\section{Summary and Results}

The mysterious content named as dark energy (DE) occupying the major
part of universe is significant in the description of cosmic
speed-up. The modified gravity theories are assumed to be effective
in understanding cosmic acceleration by induction of so-called dark
matter components in the form of higher order curvature invariants.
Among such theories, $f(R,T)$ represents non-minimal coupling of
matter and geometry. It provides an alternate to incorporate dark
energy components and cosmic acceleration \cite{38}. Thus
consideration of $f(R,T)$ for dynamical analysis is worthwhile,
covering the impact of higher order curvature terms and trace of
energy momentum tensor $T$. This manuscript is based on the role of
viable $f(R,T)$ model in establishment of instability range of
spherically symmetric star.

Our exploration regarding viability of the $f(R,T)$ model reveals
that the selection of $f(R,T)$ model for dynamical analysis is
constrained to the form $f(R,T)=f(R)+\lambda T$, where $\lambda$ is
arbitrary positive constant. The restriction on $f(R,T)$ form
originates from the complexities of non-linear terms of trace in
analytical formulation of field equations. The $f(R, T)$ form we
have chosen mainly is $f(R,T)=R+\alpha R^2+\lambda T$, in agreement
with the stable stellar configuration and satisfies the cosmic
viability. The matter configuration is assumed to have unequal
stresses i.e., anisotropic with central vacuum cavity evolving under
expansion-free condition. Zero expansion condition on anisotropic
background reveals the significance of energy density profile and
pressure inhomogeneity in structure formation and evolution.

The field equations framed in $f(R,T)$ gravity are formulated and
their conservation is considered to study the evolution.
Conservation laws yield dynamical equations that are significant in
formation of collapse equation. In order to examine the variation
from static equilibrium, we introduced linear perturbation to all
physical parameters. The expressions for perturbed configuration of
field equations reveal expressions for energy density $\bar{\rho}$
and tangential pressure $\bar{p_{\perp}}$. The second law of
thermodynamics relating radial pressure and density with the help of
adiabatic index $\Gamma$ is considered to extract $\bar{p_{r}}$.

On account of zero expansion it is found that fluid evolution is
independent of $\Gamma$, rather instability range depends on higher
order curvature corrections and static pressure anisotropy.
Recently, the dynamical analysis of isotropic and anisotropic
spherical stars in $f(R,T)$ has been studied in \cite{27, 27a}. It
is found that perturbed form of dark source terms of collapse
equation also has the contribution of trace $T$, affecting the
stability range. Thus non-minimal coupling of the higher order
curvature terms and trace of energy momentum tensor imply a wider
range of stability, however, the fluid evolving with zero expansion
might cause drastic and unexpected variations. As expansion-free
condition produces shear blow-up in gravitating system, so it is
very captivating to extend this work for shearing expansion free
case. The results are in accordance with \cite{30} for vanishing
$\lambda$, for vanishing $\alpha$ and $\lambda$ corrections to GR
solution can be found.

In addition to model (\ref{m}), nature of various $f(R,T)$ models
i.e., $f(R,T)=R+\alpha R^n+\lambda T$, $f(R,T)=R+\alpha
R^2+\frac{\mu^4}{R}+\lambda T$ and $f(R,T)=R+\frac{\mu^4}{R}+\lambda
T$ has been briefly discussed in this section, as follows:

\begin{itemize}
\item \textbf{$f(R,T)=R+\alpha R^n+\lambda T$: }
The model $f(R,T)=R+\alpha R^n+\lambda T$ is viable for any $n\geq2$
and positive constants $\alpha$ and $\lambda$. The collapse equation
for such model with zero expansion is of the form
\begin{eqnarray}\nonumber&&
\lambda
\bar{\rho}'+\lambda_1\left\{\bar{p_r}'+(\bar{\rho}+\bar{p_r})\frac{W'_0}{W_0}+
\bar{p_r}\left(\frac{2}{r}-\frac{\alpha
n(n-1)R_0^{n-2}R'_0}{1+\alpha
nR_0^{n-1}}\right)-\frac{2}{r}\bar{p_\perp}\right\}\\\nonumber&&+
D\left[\lambda_1(\rho_0+p_{r0})\frac{w'}{W_0}+\lambda_1\frac{2}{r}
(p_{r0}-p_{\perp0})-(\lambda+\lambda_1p_{r0})\frac{\alpha
n(n-1)(R_0^{n-2}e)'}{1+\alpha
nR_0^{n-1}}\right]\\\label{51}&&+Z_{3p},
\end{eqnarray}
where pressure stresses $\bar{p_r},\bar{p_\perp}$ can be generated
from perturbed field equations and perturbed energy density is
\begin{equation}\label{52}
\bar{\rho}=-\left[\frac{\alpha n(n-1)(R_0^{n-2}e)}{1+\alpha
nR_0^{n-1}}\rho_0+\lambda_1
(p_{r0}-p_{\perp0})\frac{x}{X_0}+Z_{4p}\right]D,
\end{equation}
where $Z_{3p}$ and $Z_{4p}$ depict the perturbed dark source terms.
The N and pN limit of this model reveals that the term $\bar{p_{r}}$
and $\rho_0\frac{W'_0}{W_0}$ belong to ppN limit and so do not
contribute in evolution. In Newtonian limit the physical quantities
must satisfy the following conditions
\begin{eqnarray}\nonumber&&
p_{r0}>p_{\perp0}, \quad \alpha^2 r n(n-1)R_0^{n-2} R_0'<1+n\alpha
R_0^{n-1}, \\\nonumber&& \frac{n\alpha n(n-1)R_0^{n-2}
R'_0}{1+n\alpha R_0^{n-1}}>\rho_0'-e'.
\end{eqnarray}
The constraints on physical quantities in pN regime are
\begin{eqnarray}\nonumber&&
\quad r>2m_0, \quad \frac{r}{r+m_0}(xn(n-1)R_0^{n-1}R'_0
+\frac{2em_0}{r})<e'+\lambda,\\\nonumber&& 2\alpha e-(1+n\alpha
R_0^{n-1})\frac{\bar{y}}{r}>\frac{(r^2-m_0^2)^2}
{r^4}\left\{\frac{er^2}{(r+m_0)}\right\}', \\\nonumber&&
(r-2m_0)>\frac{(n-1)R_0^{n-1}R_0'}{R_0^{n-1}}, p_{r0}>p_{\perp0}.
\end{eqnarray}
The dynamical analysis of various models involving higher order
curvature terms, combined with the trace of energy momentum tensor
can be presented for $n\geq2$.

\item \textbf{$f(R,T)=R+\alpha R^2+\frac{\mu^4}{R}+\lambda T$ and
$f(R,T)=R+\frac{\mu^4}{R}+\lambda T$:} The perturbed form of Bianchi
identity for $f(R,T)=R+\alpha R^2+\frac{\mu^4}{R}+\lambda T$, where
$\mu$ is arbitrary constant leads to the expression for $\bar{\rho}$
as follows
\begin{equation}\label{53}
\bar{\rho}=-\left[\frac{2e\rho_0}{1+2\alpha
R_0-\mu^4R_0^{-2}}+\lambda_1\left\{\frac{x}{X_0}(p_{r0}-p_{\perp0})\right\}+(1+2\alpha
R_0-\mu^4R_0^{-2})Z_{1p}\right]D.
\end{equation}
The evolution equation becomes
\begin{eqnarray}\nonumber
&&\lambda_1\left\{\bar{p_r}'+\bar{\rho}\frac{W'_0}{W_0}
+\bar{p_r}\left(\frac{W'_0}{W_0}+\frac{2}{r}-2\frac{\alpha
R'_0+\mu^4R_0^{-3}R'_0}{1 +2\alpha
R_0-\mu^4R_0^{-2}}\right)-\frac{2\bar{p_\perp}}{r}\right\}
+D\left[\lambda_1[(\rho_0\right.\\\nonumber
&&\left.+p_{r0})(\frac{w}{W_0})'
-2(p_{r0}+p_{\perp0})(\frac{\bar{y}}{r})']+\lambda\bar{\rho'}
-\frac{2\alpha+\mu^4R_0^{-3}}{1+2\alpha
R_0-\mu^4R_0^{-2}}\left\{\lambda_1\left(p'_{r0}
+\rho_0\frac{W'_0}{W_0}
\right.\right.\right.\\\nonumber&&\left.\left.\left.
+p_{r0}\left(\frac{2}{r}+\frac{W'_0}{W_0}-2\frac{\alpha
R'_0+\mu^4R_0^{-3}R'_0}{1 +2\alpha
R_0-\mu^4R_0^{-2}}\right)\right)\right\}
+\lambda\left(e'+e[\rho'_0-\frac{2\alpha R'_0}{1+2\alpha
R_0}]\right) \right.\\\label{54}&&\left.+(1+2\alpha
R_0-\mu^4R_0^{-2})Z_{5p}\right]=0.
\end{eqnarray}
$Z_{5p}$ denote the perturbed dark source entries. The N and pN
limits are obtained by avoiding the terms lying in ppN region. To
maintain the viability of model in Newtonian era, following
inequalities must hold
\begin{equation}\nonumber
\quad \alpha R'_0+\mu^4R_0^{-3}R'_0<1+2\alpha R_0-\mu^4R_0^{-2},
\quad \frac{\alpha R'_0+1+2\alpha R_0-\mu^4R_0^{-2}R'_0}{1+2\alpha
R_0-\mu^4R_0^{-2}}>\rho_0'-e'.
\end{equation}
In pN regime the system remains stable as long as following ordering
relations are satisfied.
\begin{eqnarray}\nonumber&&
\quad \frac{r}{r+m_0}(\alpha R'_0+\mu^4R_0^{-3}R'_0)<e'+\lambda
T_0,\\\nonumber&& 2\alpha e-\mu^4R_0^{-2}-(1+2\alpha
R_0-\mu^4R_0^{-2})\frac{\bar{y}}{r}>\frac{(r^2-2m_0^2)^2}
{r^4}\left\{\frac{er^2}{(r+m_0)}\right\}'.
\end{eqnarray}
The collapse equation for $f(R,T)=R+\frac{\mu^4}{R}+\lambda T$ can
be obtained by setting $\alpha=0$ in Eq. (\ref{54}), likewise the
restrictions on physical quantities can be found.
\end{itemize}

\section{Appendix}

\begin{eqnarray}\setcounter{equation}{1}\nonumber
Z_1(r,t)&=&\left[\left\{\frac{1}{f_R
W^2}\left(\frac{f-Rf_R}{2}
-\frac{\dot{f_R}}{W^2}\Theta
-\frac{f'_R}{X^2}\left(\frac{X'}{X}
-\frac{2Y'}{Y}\right)+\frac{f''_R}{X^2}\right)\right\}_{,0}\right.\\\nonumber
&&\left.+\left\{\frac{1}{f_RW^2X^2}
\left(\dot{f'_R}-\frac{W'}{W}\dot{f_R}
-\frac{\dot{X}}{X}f_R'\right)\right\}_{,1}\right]f_R W^2
-\left\{\left(\frac{\dot{X}}{X}\right)^2\right.\\\nonumber
&&\left.+2\left(\frac{\dot{Y}}{Y}\right)^2
+\frac{3\dot{W}}{W}\Theta
\right\}\frac{\dot{f_R}}{W^2}
+\frac{\ddot{f_R}}{W^2}\Theta-\frac{2f'_R}{X^2}\left\{\frac{\dot{W}}{W}\left(\frac{X'}{X}
-\frac{Y'}{Y}\right)\right.\\\nonumber
&&\left.+\frac{\dot{X}}{X}\left(\frac{2W'}{W}
+\frac{X'}{X}+\frac{Y'}{Y}\right)+
\frac{\dot{Y}}{Y}\left(\frac{W'}{W}-\frac{3Y'}{Y}\right)\right\}+\frac{\dot{W}}{W}(f-Rf_R)
\\\nonumber
&&
+\frac{f_R''}{X^2}\left(\frac{2\dot{W}}{W}+\frac{\dot{X}}{X}\right)
+\frac{1}{X^2}\left(\dot{f'_R}-\frac{W'}{W}\dot{f_R}
\right)\left(\frac{3W'}{W}
+\frac{X'}{X}+\frac{2Y'}{Y}\right),\\\label{B3}
\\\nonumber
Z_2(r,t)&=&\left[\left\{\frac{1}{f_RW^2X^2}\left(\dot{f_R}'-\frac{W'}{W}\dot{f_R}
-\frac{\dot{X}}{X}f_R'\right)\right\}_{,0}+\left\{\frac{1}{f_RX^2}\left(\frac{Rf_R-f}{2}
\right.\right.\right.
\\\nonumber &&\left.\left.\left.-\frac{\dot{f_R}}{W^2}\left(\frac{\dot{W}}{W}
-\frac{2\dot{Y}}{Y}\right)-\frac{f'_R}{X^2}\left(\frac{W'}{W}
+\frac{2Y'}{Y}\right)+\frac{\ddot{f_R}}{W^2}\right)\right\}_{,1}
\right]f_RX^2-
\\\nonumber
&&\frac{\dot{f_R}}{W^2}\left\{\frac{W'}{W}\left(\frac{\dot{W}}{W}+\frac{\dot{X}}{X}\right)
+\frac{X'}{X}\left(\frac{\dot{W}}{W}-\frac{2\dot{Y}}{Y}\right)
+\frac{2Y'}{Y}\left(\frac{\dot{X}}{X}-\frac{\dot{Y}}{Y}\right)\right\}
\\\nonumber
&&+(Rf_R-f)\frac{X'}{X}-\frac{1}{W^2}\left(\frac{\dot{W}}{W}+\frac{3\dot{X}}{X}
+\frac{2\dot{Y}}{Y}\right)\left(\frac{W'}{W}\dot{f_R}
+\frac{\dot{X}}{X}f_R'\right.\\\nonumber
&&\left.-\dot{f_R}'\right)-\frac{f'_R}{X^2}\left\{\frac{W'}{W}\left(\frac{W'}{W}+\frac{3X'}{X}\right)+
\frac{2Y'}{Y}\left(\frac{3X'}{X}
+\frac{Y'}{Y}\right)\right\}+\frac{\ddot{f_R}}{W^2}
\\\label{B4} &&\times\left(\frac{W'}{W}
+\frac{2X'}{X}\right)
+\frac{f''_R}{X^2}\left(\frac{W'}{W}
+\frac{2Y'}{Y}\right).
\end{eqnarray}
\begin{eqnarray}\nonumber
Z_{1p}&=& 2\alpha W_0^2\left[\frac{1}{W_0^2X_0^2(1+2\alpha R_0)}\left\{e'
-e\frac{W'_0}{W_0}-\frac{x}{X_0}R'_0\right\}\right]_{,1}
+\frac{1}{1+2\alpha R_0}\left[e
\right.\\\nonumber && \left.-[\lambda T_0-\alpha R_0^2]\left(\frac{w}{W_0}+\frac{e}{1+2\alpha R_0}\right)
-\frac{2\alpha}{X_0^2}\left\{\left(\frac{X'_0}{X_0}-\frac{2}{r}\right)\left(
2R'_0\left(\frac{w}{W_0}
\right.\right.\right.\right.\\\nonumber & &\left.\left.\left.\left.
+\frac{x}{X_0}\right)-e'-\frac{2 \alpha e}{1+2\alpha R_0}R'_0\right)
+ R''_0 \left(\frac{2w}{W_0}+\frac{x}{X_0}\right)-2R'_0\left(\frac{x}{X_0}\left(\frac{2W'_0}{W_0}
\right.\right.\right.\right.\\\nonumber
 & &\left.\left.\left.\left.+
\frac{X'_0}{X_0}+\frac{1}{r}\right)+\frac{\bar{y}}{r}\left(\frac{W'_0}{W_0}-\frac{3}{r}\right)\right)
+\left(e'-e\frac{W'_0}{W_0}\right)\left(\frac{3W'_0}{W_0}+
\frac{X'_0}{X_0}\right.\right.\right.\\\label{Z1p}
 & &\left.\left.\left.+\frac{2}{r}\right)\right\}\right]
\\\nonumber
Z_{2p}&=&X_0^2(1+2\alpha R_0)\left[\frac{1}{X_0^2(1+2\alpha R_0)}\left\{e+\frac{2\alpha}{X_0^2}\left\{
\left(\frac{W'_0}{W_0}+\frac{2}{r}\right)\left([\frac{2 \alpha e}{1+2\alpha R_0}
\right.\right.\right.\right.\\\nonumber & &\left.\left.\left.\left.
+\frac{4x}{X_0}]R'_0-e\right)
-R'_0[\left(\frac{w}{W_0}\right)'
+\left(\frac{\bar{y}}{r}\right)']\right\}+[\lambda T_0
-\alpha R_0^2]\left(\frac{e}{1+2\alpha R_0}
\right.\right.\right.\\\nonumber &&\left.\left.\left.+\frac{x}{X_0}\right)\right\}\right]_{,1}
+xX_0(1+2\alpha R_0)\left[\frac{-1}{X_0^2(1+2\alpha R_0)}\left\{\frac{4\alpha R'_0}{X_0^2}\left(\frac{W'_0}{W_0}+\frac{2}{r}\right)
\right.\right.\\\nonumber &&\left.\left.
+\alpha R_0^2-\lambda T_0\right\}\right]_{,1}
+\frac{2\alpha}{X_0^2}\left[R''_0\left\{\left(\frac{w}{W_0}\right)'-2\left(\frac{W'_0}{W_0}
+\frac{2}{r}\right)\left(\frac{e}{1+2\alpha R_0}
\right.\right.\right.\\\nonumber &&\left.\left.\left.+\frac{x}{X_0}\right)+\left(\frac{\bar{y}}{r}\right)'
\right\}-R'_0\left\{\frac{W'_0}{W_0}\left[\left(\frac{2w}{W_0}\right)'+\left(\frac{3x}{X_0}\right)'
\right]
+3\frac{X'_0}{X_0}\left[\left(\frac{x}{X_0}\right)'
\right.\right.\right.\\\nonumber & &\left.\left.\left.
+2\left(\frac{\bar{y}}{r}\right)'\right]
+\frac{2}{r}\left[\left(3\frac{x}{X_0}\right)'+2\left(\frac{\bar{c}}{r}\right)'
\right]\right\}+\left(\frac{2x}{X_0}R'_0
-e\right)
\left\{\left(\frac{W'_0}{W_0}\right)^2
\right.\right.\\\nonumber & &\left.\left.3\frac{X'_0}{X_0}\left(\frac{W'_0}{W_0}+\frac{2}{r}\right)+\frac{2}{r^2}\right\}\right]
+e\frac{X'_0}{X_0}-[\lambda T_0
-\alpha R_0^2]\left(\frac{2e}{1+2\alpha R_0}\frac{X'_0}{X_0}
\right.\\\label{Z2p} & &\left.+\frac{x}{X_0}\right)
\end{eqnarray}
\begin{eqnarray}\nonumber
Z_{3}&=&\frac{1+2\alpha R_0}{X_0^2}\left[\frac{w''}{W_0}+\frac{\bar{y}''}{r}-\frac{W''_0}{W_0}
\left(\frac{w}{W_0}+\frac{2x}{X_0}\right)+\frac{W'_0}{W_0}\left\{\frac{2x}{X_0}
\left(\frac{X'_0}{X_0}
-\frac{1}{r}\right)\right.\right.\\\nonumber & &\left.\left. +\left(\frac{\bar{y}}{r}\right)'
-\left(\frac{x}{X_0}\right)'\right\}
+\frac{X'_0}{X_0}\left\{\frac{2xX_0'}{rX_0}-\left(\frac{w}{W_0}\right)'-\left(\frac{\bar{y}}{r}\right)'\right\}+
\left\{\left(\frac{w}{W_0}\right)'\right.\right.\\\nonumber & &\left.\left.
-\left(\frac{x}{X_0}\right)'\right\}\frac{1}{r}
\right]-\frac{2\alpha e}{1+2\alpha R_0}\left\{\frac{\lambda T_0
-\alpha R_0^2}{2}-\frac{2\alpha}{X_0^2}\left(R'_0\left(\frac{W'_0}{W_0}-\frac{X'_0}{X_0}
\right.\right.\right.\\\nonumber & &\left.\left.\left.+\frac{1}{r}\right)-R''_0\right)\right\}
-\frac{2\alpha}{X_0^2}\left\{e''+\frac{2x}{X_0}R''_0+\left(\frac{W'_0}{W_0}-\frac{X'_0}{X_0}
+\frac{1}{r}\right)\left(\frac{2x}{X_0}R'_0\right.\right.\\\label{Z3} & &\left.\left.
-e'\right)\right\}
\end{eqnarray}

\textbf{Acknowledgment}

I.N. acknowledges Dr. Hafiza Rizwana Kausar for her support and fruitful discussions.

\end{document}